\documentclass[5p]{elsarticle}
\journal{Computer Networks}

\usepackage{amsmath,amssymb,amsfonts,dsfont}
\usepackage{graphicx}
\usepackage{textcomp}
\usepackage[dvipsnames]{xcolor}
\usepackage{subfigure,dblfloatfix}
\usepackage{tabularx}
\usepackage{algorithm}
\usepackage[noend]{algpseudocode}
\usepackage[hyphens]{url}

\newcommand{\Fig}[1]{Figure~\ref{fig:#1}}

\newcommand{\Sec}[1]{Sec.~\ref{sec:#1}}

\newcommand{\Eq}[1]{(\ref{eq:#1})}
\newcommand{\Alg}[1]{Alg.~\ref{alg:#1}}
\newcommand{\Line}[1]{Line~\ref{line:#1}}
\newcommand{\ind}[1]{\mathds{1}_{[#1]}}

\input{my_math.sty}
\newcommand{\Mbs}{M_{\rm BS}}
\newcommand{\Mue}{M_{\rm UE}}
\newcommand{\Mmn}{M_{\rm MN}}

\newtheorem{property}{Property}
\newproof{proof}{Proof}
\newcommand{\Prope}[1]{Property~\ref{prope:#1}}

\begin{document}

\begin{frontmatter}

\title{
Eavesdropping with Intelligent Reflective Surfaces:\\
Near-Optimal Configuration Cycling
}
    \author[cnr,cnit]{Francesco Malandrino\corref{cor1}}
    \ead{francesco.malandrino@cnr.it}
    \cortext[cor1]{Corresponding author.}
    \author[cnr,cnit]{Alessandro Nordio}
    \author[polito,cnr,cnit]{Carla Fabiana Chiasserini}
    \address[cnr]{CNR-IEIIT, Torino, Italy}
    \address[cnit]{CNIT, Italy}
    \address[polito]{Politecnico di Torino, Torino, Italy}
    \journal{Computer Networks}

\begin{abstract}
Intelligent reflecting
surfaces (IRSs) have several prominent advantages, including improving 
the level of wireless communication security and privacy. In this work, 
we focus on the latter aspect and introduce a strategy to counteract the presence of passive eavesdroppers 
overhearing transmissions from a base station towards legitimate users that are 
facilitated by the presence of IRSs. 
Specifically, we envision a transmission scheme that cycles across a number of IRS-to-user assignments, 
and we select them in a near-optimal fashion, thus guaranteeing both 
a high data rate and a good secrecy rate.
Unlike most of the existing works addressing passive eavesdropping, 
the strategy we envision has low complexity and is suitable for scenarios where nodes are 
equipped with a limited number of antennas. Through our performance evaluation, we highlight the trade-off
between  the legitimate users' data rate and 
secrecy rate, and how the system parameters affect such a trade-off.
\end{abstract}


\end{frontmatter}

\section{Introduction\label{sec:intro}}
It is expected that the sixth generation (6G) of mobile communications
will exploit terahertz (THz) frequencies (e.g., 
0.1–10~THz~\cite{Aluoini-mmw-thz, Akyildiz-THz}) for indoor as well as
outdoor applications.  THz communications can indeed offer 
very high data rates, although over short 
distances due to harsh propagation conditions and severe path loss.
To circumvent these problems, massive multiple-input-multiple-output (mMIMO) communication and
beamforming techniques can be exploited to concentrate the transmitted
power towards the intended receiver. Further, the use of intelligent reflecting
surfaces (IRSs)~\cite{Wu} has emerged as a way to enable smart radio environments (SREs)~\cite{DiRenzo}.
In such works, the high-level goal is to optimize the performance, and such a goal is pursued by 
controlling and adapting the radio environment to the communication between a transmitter and a receiver.

IRSs are passive beamforming devices, composed of a large number of
low-cost antennas that receive signals from sources, customize them
by basic operations, and then forward them along the desired
directions~\cite{Liaskos,Liaskos-3,Alsharif2020}.
They have been successfully used to enhance the {\em security} of the network -- 
typically, against eavesdroppers -- as well as to improve the network performance.

As better discussed in \Sec{relwork}, existing works about IRS-based security mostly aim at optimizing 
the IRSs rotation and 
phase shift~\cite{Liaskos,Liaskos-3,Alsharif2020,dong2020secure,bereyhi2020secure,mensi2021physical,wijewardena2021physical} 
to find {\em one} high-quality configuration that guarantees both high performance and good privacy levels.  
Techniques used to this end include iterative algorithms~\cite{wijewardena2021physical}, 
particle-swarm optimization~\cite{qi2023irs}, and Kuhn-Munkres algorithms maximizing the sum-rate~\cite{sun2022optimization}. 
As in other fields, deep reinforcement learning is another very popular approach used to address the above aspects, 
as exemplified by~\cite{zhang2022deep}.

In this work, we investigate the secrecy performance of IRS-based communications, considering 
the presence of a malicious receiver passively overhearing the downlink transmission intended 
for a legitimate user.
Our high-level strategy is predicated upon (i) identifying a small number of {\em configurations}, i.e., 
IRS-to-user equipment (UE) assignments that guarantee both high data rate and good secrecy rate, and (ii) cycling among such 
configurations.
It is worth pointing out that the strategy we propose does not aim at generically optimizing the IRSs' rotation and phase shifts, as done in the literature; rather, IRSs are always oriented towards one of the UEs, and we have to choose {\em which} one. Operating this simplification greatly 
reduces the solution space to explore, at a negligible cost in performance.
Furthermore, while most existing solutions look for {\em one} high-quality configuration, 
we select a near-optimal {\em set of configurations} among which to cycle, to further enhance the robustness of the system.

Therefore, our main contributions can be summarized as follows:
\begin{itemize}
    \item we propose a new approach to IRS-based communications in the presence of an eavesdropper, 
    predicated upon periodically switching among multiple IRS-to-UE assignments (configurations);
    \item we formulate the problem of selecting a near-optimal set of said configurations, 
    balancing the -- potentially conflicting -- requirements of high secrecy rate and high data rate;
    \item after proving its NP (non-polynomial)-hardness , we solve such a problem through an iterative scheme 
    called {\sf ParallelSlide}, yielding near-optimal solutions with a polynomial computational 
    complexity.
\end{itemize}

The remainder of the paper is organized as follows. After discussing related work in \Sec{relwork},
\Sec{comms} describes the physical-layer aspects of the IRS-based communication we study.
\Sec{problem} describes how the communication from a base station to a set of legitimate users is 
facilitated by IRSs, 
and details how switching between different IRS-to-UE assignments can take place.
In that context, \Sec{algo} first presents the problem we solve when selecting the best 
configurations and characterizes its complexity; 
then it introduces our {\sf ParallelSlide} solution and proves its properties. 
\Sec{peva} characterizes the trade-offs 
our approach is able to attain, and it compares the performance of the proposed solution against 
state-of-the-art alternatives. Finally, we conclude the paper in
\Sec{concl}.

\begin{figure} 
\includegraphics[width=0.9\columnwidth]{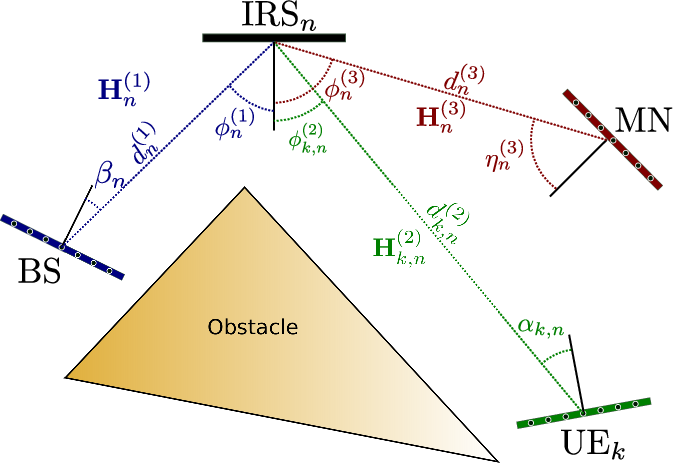}
\centering
\caption{\label{fig:model} Communication model: a base station (BS) is
  transmitting toward the $k$-th UE, thanks to the help of
  the $n$-th IRS. The LoS link between BS and UE is blocked by an obstacle.
 The $k$-th  UE  is the victim of the malicious node (MN), which
  intercepts the signals reflected by the IRSs.}
\end{figure}

\section{Related Work}
\label{sec:relwork}

The main research area our work is related to is physical-layer security for IRS-assisted wireless networks.

As discussed in~\cite{zhou2020enabling}, IRSs can be efficiently used
to improve the security and privacy of wireless communications,
as they can make the channel better for legitimate users,
and worse for malicious ones.
As an example, the authors of~\cite{zhou2020user} targeted the case of {\em aligned} eavesdroppers, 
lying between the transmitter and the legitimate receiver: in this case, the authors 
envisioned avoiding direct transmissions, and using IRSs to maximize the secrecy rate. 

Jamming is an effective, even if harsh, method to improve privacy by making the eavesdropper's 
channel worse: as an example, \cite{guan2020intelligent}~envisioned using IRSs to both serve 
legitimate users and jam the malicious one, maximizing the secrecy rate subject to power constraints.
In MIMO scenarios, passive eavesdroppers can be blinded through standard beamforming techniques, 
thanks to the so-called secrecy-for-free property of MIMO systems with large antenna arrays. 
Several recent works, including~\cite{bereyhi2020secure}, aimed at achieving the same security level 
against active attackers, by leveraging filtering techniques and the fact that legitimate 
and malicious nodes are statistically distinguishable from each other. 
In a similar scenario, \cite{dong2020secure} presented an alternating optimization that jointly 
optimizes both transmitter and IRS parameters in order to maximize the secrecy rate.
The authors of~\cite{mensi2021physical} addressed a vehicular scenario, finding that IRSs are more effective 
than leveraging vehicular relays to attain physical-layer security. 

Many works focussed on the problem of configuring the IRSs available in a given scenario 
to optimize one or more target metrics, e.g., performance, secrecy rate, or cost. 
Examples include~\cite{wijewardena2021physical}, where an iterative algorithm is used to 
optimize the sum-rate. In a similar setting and with the same objective, \cite{qi2023irs}~resorted 
to particle-swarm optimization, owing to the problem complexity and to the need for quick convergence; 
for similar reasons, the authors of~\cite{zhang2022deep} leveraged deep reinforcement learning. 
An unusual twist is represented by~\cite{sun2022optimization}, which focused on visible-light communication 
and optimizes the configuration of IRSs (i.e., mirrors) to optimize the sum-rate, through a Kuhn-Munkres algorithm.

All the works we have discussed share two very important features, to wit:
\begin{itemize}
    \item they seek to find {\em one} high-quality IRS configuration, and
    \item they build such a configuration from the ground up, i.e., optimizing 
    the phase-shifting vectors of the individual IRS elements.
\end{itemize}
We depart from such features by (i) selecting a set of {\em multiple} configurations among which to cycle, and (ii) 
confine ourselves to configurations where each IRS points to a user, on the grounds that such configurations are very 
likely to be the most useful, and restricting our attention to them significantly decreases the complexity.

\section{Communication Model}
\label{sec:comms}

We consider a wireless network operating in the THz bands, composed of: 
\begin{itemize}
\item A base station (BS) equipped with a uniform linear array (ULA) of antennas composed of $\Mbs$~isotropic 
elements and transmitting $K$ data streams, one for each legitimate
  user;
\item $K$ legitimate users (UEs), each equipped with an ULA composed of $\Mue$
  isotropic antenna elements;
\item $N$ IRSs ($N\ge K$) composed of arrays of elements (or meta-atoms) arranged in a square grid.
The IRSs contribute to the BS-UEs communication
  by appropriately forwarding the BS signal toward the users. 
  Notice that, given $K$ legitimate UEs, at any time instant only $K$ IRSs are used.
\item A passive eavesdropper (or malicious node, MN), whose ULA is composed of
  $\Mmn$ isotropic antenna elements. The goal of the MN is to eavesdrop the communication from 
  the BS towards one of the $K$ legitimate UEs, by intercepting the signals reflected by the IRSs. 
To do so, the MN exploits the directivity provided by its ULA by pointing
  it towards the IRS serving the UE that the MN wants to eavesdrop. 
\end{itemize}

We assume ULA made of isotropic elements whose gain is
  0\,dBi.  In practice, each element of the ULA is driven 
  by a phase shifter; thus,  by adjusting the phase relationship between
  the antenna elements, the ULA radiation pattern can be
  manipulated to generate and steer the beam in a specific direction,
  or change its shape. Interestingly, in  next-generation telecommunication and 
  radar systems, it is envisioned that phased arrays are replaced with transmitarrays, i.e., 
  high-gain antenna systems manufactured with multi-layer printed
  circuit technology (usually on low-loss substrates as, e.g., quartz or
  silicon) designed for applications in the 10–300\,GHz frequency
  range.

In the following, boldface uppercase and lowercase letters denote
matrices and column vectors, respectively, while uppercase
calligraphic letters are used for sets. $\Id_k$ is the $k\times k$
identity matrix. For any matrix $\Am$, its transpose and conjugate
transpose are denoted by $\Am\Tran$ and $\Am\Herm$, respectively,
while $[\Am]_{i,j}$ is its $(i,j)$-th element. Finally, the symbol
$\otimes$ denotes the Kronecker product.

Below, we characterize the main elements of the system, 
namely, the channel and the IRSs (\Sec{sub-channel}), as well as the other network nodes and 
their behavior (\Sec{sub-nodes}).
To this end, we initially assume  that the position of the eavesdropper 
(hence, the channel conditions it experiences) is known, so that we can characterize the 
system performance in a clear manner.
Importantly, as also remarked later, 
this is {\em not} an assumption required by our algorithm or solution concept 
and will be dropped in the following sections.

\subsection{Channel model and IRS characterization\label{sec:sub-channel}}
We assume that no line-of-sight (LoS) path exists between the BS and
the UEs. However, communication is made possible by the ability of the
IRSs to reflect the BS signal towards the
users~\cite{noi-twc22}. Instead, the BS--IRS and IRS-UE links are
LoS as well as the IRS--MN link, as depicted
in Fig.~\ref{fig:model}. Also, the BS, all IRSs and user nodes, including
  the MN, are assumed to have the same height above ground. This
  assumption allows simplifying the discussion and the notation while
  capturing the key aspects of the system. In the following, we detail
  the  channel model and the IRS configuration.

{\bf Communication channel.} While in many works dealing with
communications on GHz bands, the channel connecting two multi-antenna
devices is often modeled according to Rayleigh or Rice distributions,
in the THz bands the channel statistic has not yet been completely
characterized. Moreover, at such high frequencies, the signal suffers
from strong free-space attenuation, and it is blocked even by small
solid obstacles.  In practice, the receiver needs to be in LoS with
the transmitter to be able to communicate.  Also, recent
studies~\cite{Xing} have highlighted that already at sub-THz
frequencies all scattered and diffraction effects can be
neglected. Multipath, if present, is due to reflection on very large
objects which, however, entails severe reflection losses. As an
example, a plasterboard wall has a reflection coefficient of about
-10\,dB for most of the incident angles~\cite{reflection}.  In
practice, the number of paths is typically very small and even reduces
to one, i.e., the LoS component, when large high-gain antennas are
used~\cite{Akyildiz2018}.  This situation occurs when the transmitter
and the receiver adopt massive beamforming techniques that generate
beams with very small beamwidth in order to concentrate the signal
energy along a specific direction and compensate for high path losses.
In such conditions, non-LoS (NLos) paths are very unlikely to show.

 Beamforming clearly improves the security of
  communication since a malicious user must be located within the beam
  cone to be able to eavesdrop the signal. In addition, 
  spatial diversity and beam configuration switching
  can make (on average)  the  channel between the BS and the eavesdropper and the one between the 
  legitimate user and the BS differ, thus further improving security.

  In this work, we denote
with $M_{\rm tx}$ and $M_{\rm rx}$  the number of antennas at the
transmitter and at the receiver, respectively, the $M_{\rm rx}\times M_{\rm tx}$
  channel matrix between any two devices can be modeled as:
\begin{equation}
  \label{eq:channel_model}
  \Hm^{\rm LOS} = a g \pv \qv\Herm \,.
\end{equation}
In \Eq{channel_model},  scalar $a$ takes
into account large-scale fading effects
due to, e.g., obstacles
temporarily crossing the LoS path between transmitter and
receiver, while coefficient $g$ accounts for the attenuation and
phase rotation due to propagation.  
More specifically,  let $d$ be the
distance between the transmitting and the receiving device and  
$\lambda$  the signal wavelength. For the BS-IRS (IRS-UE) link, we denote with $G$ 
be the array gain of the transmitter (receiver) and with  $S$ the effective area of the receiver (transmitter). 
Then the expression for $g$ is given by:
     \begin{equation}\label{eq:c_Area}
      g = \sqrt{\frac{G S}{4\pi d^2}}\ee^{\jj \frac{2 \pi}{\lambda}d } \,.
    \end{equation}

 Finally,  vector $\pv$ of size $M_{\rm rx}$ and 
  vector $\qv$ of size $M_{\rm tx}$ are norm-1 and represent 
 the spatial signatures of, respectively, the receive and the transmit
antenna array.  The spatial signature of an ULA composed of $M_{\rm
  z}$ ($z\in\{{\rm BS},{\rm UE},{\rm MN}\}$) isotropic elements, spaced
by $\lambda/2$ and observed from an angle $\beta$ (measured with
respect to a direction orthogonal to the ULA), is given by the
$M_z$-size vector $\sv(\beta,M_z)$, whose $m$-th element is given by:
\begin{equation}
  \label{eq:signature}
  [\sv(\beta,M_z)]_m = \frac{1}{\sqrt{M_z}}\ee^{-\jj\frac{\pi}{2} (M_z-1)\sin \beta}\ee^{-\jj \pi (m-1)\sin \beta}\,.
\end{equation}
This relation applies to devices equipped with ULAs such as the BS,
the UEs, and the MN. However, it can also be applied to IRSs with elements 
spaced by $\lambda/2$, since
their planar configuration can be viewed as a superposition of several
ULAs.

{\bf IRS characterization.} IRSs are made of meta-atoms (modeled as
elementary spherical scatterers) whose scattered electromagnetic field
holds in the far-field regime~\cite{direnzo2,Ozdogan}.  
  We assume that the meta-atoms can reflect the impinging 
  signal without significant losses and apply to it a (controlled) continuous
  phase shift, which is independent of the frequency.  Such IRS model,
  although ideal, has  been widely used \cite{twc22} and holds with
  a fairly good approximation if the transmitted signal lies in the
  IRS operational bandwidth which, in the most common designs,
  amounts to about 10-15\% of the central frequency.

The $n$-th IRS, $n=1,\ldots,N$, is composed of $L_n^2$
meta-atoms arranged in an $L_n\times L_n$ square grid and spaced by $\lambda/2$.  
Thus, the area of the $n$-th IRS is given by $A_n=L_n^2\lambda^2/4$.
The meta-atom at position $(\ell,\ell')$
in the $n$-th IRS applies a phase-shift $\theta_{n,\ell,\ell'}$, to
the signal impinging on it.  In many works that assume rich scattering
communication channels such phase-shifts are independently optimized
in order to maximize some performance figures. However, under the
channel model in~\Eq{channel_model}, phase-shifts are related to each
other according to the linear equation~\cite{scattermimo, Wu-Zhang}:
\begin{equation}\label{eq:phase_shift}
  \theta_{n,\ell,\ell'} = \pi q_n\left(\ell-1-\frac{L_n-1}{2}\right) + \psi_n
\end{equation}
where $q_n$ and $\psi_n$ control, respectively, the direction and the
phase of the reflected signal. For simplicity, we can arrange the phase
shifts $\theta_{n,\ell,\ell'}$ in the diagonal matrix
\[  \bar{\Thetam}_n = \Id_{L_n}\otimes \Thetam_n \]
where $\Thetam_n=\diag(\ee^{\jj \theta_{n,1,\ell'}},\ldots,
\ee^{\jj \theta_{n,L_n,\ell'}})$. Also,  we recall that
$\otimes$ denotes the Kronecker product and  $\Id_{L_n}$ is the
identity matrix of size $L_n$. To clarify how
  IRSs are configured, consider the example depicted in
  Fig.~\ref{fig:model} where $\phi^{(1)}_n$ is the angle of arrival
  (AoA) of the BS signal on the $n$-th IRS and $\phi^{(2)}_{n,k}$ is
  the direction of the $k$-th UE as observed from the $n$-th
  IRS. Then, to let the $n$-th IRS reflect the BS signal towards
  the $k$-th UE, we set $q_n=q_{n,k}$ in~\Eq{phase_shift}
  where~\cite{noi-twc22}:
\begin{equation}
  q_{n,k} = \sin \phi^{(1)}_n - \sin \phi^{(2)}_{n,k}\,. 
  \label{eq:q_n}
\end{equation}

\begin{figure} 
\includegraphics[width=1\columnwidth]{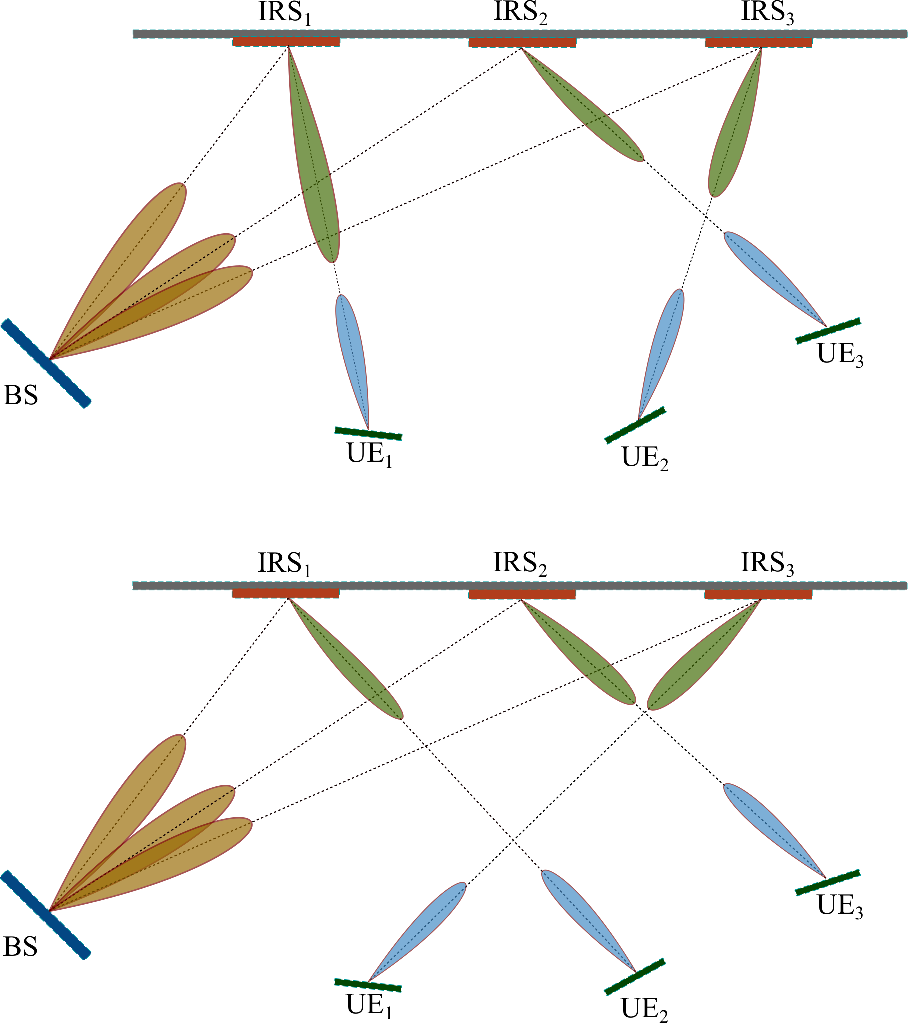}
\centering
\caption{\label{fig:configurations}Two possible configurations for a
  network with $N=K=3$. The top configuration corresponds to the
  map $\nu_a(1)=1$, $\nu_a(2)=3$, and $\nu_a(3)=2$, while the bottom configuration
  corresponds to the map $\nu_b(1)=3$, $\nu_b(2)=1$, and
  $\nu_b(3)=2$.}
\end{figure}

In a scenario with many IRSs and many UEs, where an IRS serves at most a single
UE, we can define a map, $c$, between the set of UEs and the set of IRSs, defined as
\begin{equation} 
\nonumber
c : \{1,\ldots, K\} \to \{1,\ldots,N\}. 
\end{equation}
This map, in the following referred to as {\em configuration},
specifies which UE is served by which IRS.
We also denote as~$\nu_c(k)\in\Nc$ the UE served by the $k$-th IRS under configuration~$c$.
It follows that,
if
$\nu_c(k)=n$, we mean that UE $k$ is served by  IRS $n$; in this case, 
the parameter $q_n$ of the $n$-th IRS has to take the value in~\Eq{q_n}.
With $N$ IRSs and $K$ UEs, the number
of possible configurations is $C=N!/(N-K)!$ and the corresponding set
is denoted by $\Cc$.
Under configuration $c\in\Cc$,
\begin{itemize}
  \item IRS $\nu_c(k)$ forwards the BS signal towards the $k$-th UE and, by symmetry, we assume that the $k$-th UE points its beam
    towards the $\nu_c(k)$-th IRS;
  \item we denote by $\bar{\Thetam}_{n,c}$ the matrix of the phase-shifts at the $n$-th IRS.
\end{itemize}
An example of possible configurations for a network
composed of $N=3$ IRSs and $K=3$ UEs is depicted in
Fig.~\ref{fig:configurations}.

\subsection{Behavior of BS, UEs, and MN}
\label{sec:sub-nodes}

{\bf Base station.}  The BS transmits a signal with bandwidth
$B$ and wavelength $\lambda$. Such signal contains $K$ data streams,
one for each UE.  Let~$x_k$ be the zero-mean, unit-variance Gaussian
complex i.i.d. random symbol generated for the $k$-th stream at a
given time. Also, let $\gammav_k$ be the beamforming vector 
of size $\Mbs$, employed for transmitting $x_k$.
Then, the signal transmitted by the BS is given
  by the size-$\Mbs$ vector
\begin{equation} 
\label{eq:t}
\tv= \Gammam \xv 
\end{equation}
where $\Gammam = [\gammav_1, \ldots, \gammav_K]$
  is the $\Mbs \times K$ precoding matrix and $\xv=[x_1,\ldots,
  x_K]\Tran$. We assume that the total transmit power is limited by
$P_t$, i.e., $\EE[|\tv|^2] = \|\Gammam\|_{\rm F}^2 \le P_t$, with 
$\|\cdot\|_{\rm F}$ denoting the Frobenius norm.

{\bf Legitimate receivers (UEs).} The
signal received by the $k$-th UE under the $c$-th configuration is given by~\cite{noi-twc22}
\begin{eqnarray}
  r_{k,c} &=& \underbrace{\fv_{k,c}\Herm \sum_{n=1}^N\Hm^{(2)}_{k,n}\bar{\Thetam}_{n,c}\Hm^{(1)}_n}_{\widetilde{\hv}_{n,c}\Herm}\tv +n_k\label{eq:rk}
\end{eqnarray}
where
\begin{itemize}
  \item $n_k\sim\Nc_\CC(0,N_0 B)$ is additive Gaussian complex noise  with zero-mean and variance $N_0 B$, and $N_0$ is its  power spectral density;
  \item the size-$\Mue$ vector $\fv_{k,c}$ represents the beamforming
    at the $k$-th UE under the $c$-th configuration. In particular, we
    assume that the UE's ULA is only capable of analog beamforming; 
    thus, $\fv_{k,c}=\sv(\alpha_{k,n},\Mue)$ where $n=\nu_c(k)$, i.e.,
    the radiation pattern of the $k$-th UE ULA points
    to the $\nu_c(k)$-th IRS;
\item $\Hm^{(1)}_n$ is the $L_n^2\times \Mbs$ channel matrix
  connecting the BS to the $n$-th IRS; according to the channel model
  in~\Eq{channel_model}, it is given by $\Hm^{(1)}_n =
  a_n^{(1)}g_n^{(1)}\pv_n^{(1)}{\qv_n^{(1)}}\Herm$ where
  $\qv_n^{(1)}\mathord{=}\sv(\beta_n,\Mbs)$,
  $\pv_n^{(1)}\mathord{=}\frac{1}{\sqrt{L_n}}\onev_{L_n}\otimes
  \bar{\pv}_n^{(1)}$ and $\bar{\pv}_n^{(1)}=\sv(\phi^{(1)}_n,L_n)$. 
  Moreover, $g^{(1)}_n\mathord{=}\frac{\sqrt{\Mbs A_n}}{\sqrt{4\pi}
    d^{(1)}_n}\ee^{\jj \frac{2 \pi}{\lambda}d^{(1)}_n}$ and
  $d^{(1)}_n$ is the distance between the BS and the $n$-th IRS;
\item
  $\Hm^{(2)}_{k,n}\mathord{=}a^{(2)}_{k,n}g^{(2)}_{k,n}\pv_{n,k}^{(2)}{\qv_{n,k}^{(2)}}\Herm$
  is the $\Mue \times L_n^2$ channel matrix connecting the $n$-th IRS
  to the $k$-th UE where, according to~\Eq{channel_model}, we have $\pv_{k,n}=\sv(\alpha_{k,n},\Mue)$ and
  $\qv_{n,k}^{(2)}= \frac{1}{\sqrt{L_n}}\onev_{L_n}\otimes
  \bar{\qv}_{k,n}^{(2)}$. Moreover, 
  $\bar{\qv}_{k,n}^{(2)}\mathord{=}\sv(\phi_{k,n}^{(2)},L_n)$ and
  $g^{(2)}_{k,n} = \frac{\sqrt{\Mue A_n}}{\sqrt{4\pi}
    d^{(2)}_{n,k}}\ee^{\jj \frac{2 \pi}{\lambda}d^{(2)}_{n,k}}$. Finally,
  $d^{(2)}_{n,k}$ is the distance between the $n$-th IRS and $k$-th UE.
\end{itemize}
  Notice
  that, by assuming that all network nodes have the same height over
  the ground, the $n$-th IRS can be viewed as a superposition of $L_n$
  identical ULAs. Thus, its spatial signature can be written in a 
  compact form by using the Kronecker product, as indicated above.
  Also, the angles $\beta_n$, $\phi_n^{(1)}$, $\phi_{k,n}^{(2)}$, and
  $\alpha_{k,n}$ are specified in Fig.~\ref{fig:model} and are
  measured with respect to the normal to the corresponding ULA or
  IRS.

  By collecting in vector $\rv_c=[r_{1,c},\ldots,r_{K,c}]\Tran$
the signals received by the $K$ UEs and by recalling~\Eq{t}, we
can write: 
\begin{equation}
  \rv_c = \widetilde{\Hm}_c\Herm \tv + \nv
      = \widetilde{\Hm}_c\Herm \Gammam\xv + \nv\label{eq:rv}
    \end{equation}
where $\widetilde{\Hm}_c=[\widetilde{\hv}_{1,c},\ldots,\widetilde{\hv}_{K,c}]$ and $\nv = [n_1,\ldots,n_K]\Tran$.

{\bf Malicious node.}
By eavesdropping the communication, the MN acts as an additional
receiver. When  configuration $c$ is applied and the
MN's ULA points to the $n$-th IRS, the received signal can be written similarly 
to~\Eq{rk}, as
\begin{eqnarray}\label{eq:o}
  o_{n,c}  &=& \underbrace{\bv_n\Herm \sum_{m=1}^N \Hm_m^{(3)}\bar{\Thetam}_{m,c}\Hm_m^{(1)}}_{\widetilde{\bv}_{n,c}\Herm}\tv + \zeta
\end{eqnarray}
where $\Hm_m^{(3)}=a_m^{(3)}c_m^{(3)}\pv_m^{(3)}{\qv_m^{(3)}}\Herm$ is
the $L_n^2 \times \Mmn$ channel matrix connecting the $n$-th IRS to the MN,
$\pv_m^{(3)}=\sv(\eta_m^{(3)},\Mmn)$,
$\qv_m^{(3)}=\frac{1}{\sqrt{L_m}}\onev_{L_m}\otimes
\bar{\qv}_m^{(3)}$, $\bar{\qv}_m^{(3)}=\sv(\phi^{(3)}_m,L_m)$ (see
Fig.~\ref{fig:model}).  Also,
$g_m^{(3)}=\frac{\sqrt{\Mmn A_m}}{\sqrt{4\pi} d^{(3)}_m}\ee^{\jj \frac{2
  \pi}{\lambda}d^{(3)}_m}$ where $d^{(3)}_m$ is the distance between
the MN and the $m$-th IRS. Finally,
$\zeta\sim\Nc_\CC(0,N_0B)$ represents the additive noise
at the receiver and $\bv_n=\sv(\eta_n^{(3)},\Mmn)$ is the norm-1 beamforming vector.

\section{Network Management Mechanism}
\label{sec:problem}

Under our network management mechanism, the BS and the legitimate nodes {\em switch}, periodically 
and in a synchronized manner, 
between different  configurations, i.e., IRS-to-UE assignment. So doing, 
they can counteract the eavesdropper's efforts at guessing the 
current configuration. At the same time, the BS must be careful not to use configurations 
with poor performance, i.e., yielding a low data rate. 
Let us 
denote with $R(k,c)$ the rate
experienced by user~$k$ under configuration~$c\in \Cc$, and with 
$\text{SR}(n,k,c)$ the secrecy rate obtained under
configuration~$c$ when the victim is user~$k$ and the eavesdropper is
listening to IRS~$n$.
Furthermore, let~$k^\star$ identify the
eavesdropper's victim. 

In the following, we first define the performance metrics of interest, namely, 
the data rate and the secrecy rate of  legitimate users (\Sec{sub-metrics}); then, we introduce  the 
communication scheme that is adopted by the BS, the legitimate users, and the MN (\Sec{sub-time}).

\subsection{Performance metrics}
\label{sec:sub-metrics}

The SINR achieved at each UE depends on the precoding strategy
employed at the BS, i.e., on the choice of the precoder $\Gammam$.
For example, the zero-forcing (ZF) precoder permits to remove the
inter-user interference while providing good (although not optimal)
performance. Under the $c$-th configuration, $\nu_c$, the ZF precoder
is obtained by solving for $\Gammam_c$ the equation
$\widetilde{\Hm}_c\Herm\Gammam_c=\mu\Pim$ where $\Pim=\diag(\pi_1,\ldots,
\pi_K)$ is an arbitrary positive diagonal matrix and the scalar $\mu$
should be set so as to satisfy the power constraint
$\|\Gammam_c\|_{\rm F}^2 \le P_t$. The diagonal elements of $\Pim$
specify how the transmit power is shared among users; as an example, if $\Pim$ is
proportional to the identity matrix, the same fraction of signal power is
assigned to each UE. 
The expression of the ZF precoder $\Gammam_c$ is then given by: 
\begin{equation}\label{eq:Gamma}
  \Gammam_c \triangleq \frac{\sqrt{P_t}\widetilde{\Hm}_c^+\Pim^{\frac{1}{2}}}{\|\widetilde{\Hm}_c^+\Pim^{\frac{1}{2}}\|_{\rm F}} 
\end{equation}
where $\widetilde{\Hm}_c^+=\widetilde{\Hm}_c\Herm(\widetilde{\Hm}_c\widetilde{\Hm}_c\Herm)^{-1}$
is the Moore-Penrose pseudo-inverse of $\widetilde{\Hm}_c$.  

The
SINR at the $k$-th UE is then given by: 
\begin{equation}
{\rm SINR}^{{\rm UE}}_{k,c} = \frac{P_t}{N_0B\|\widetilde{\Hm}_c^+\Pim^{\frac{1}{2}}\|^2_{\rm F}} \,.
\end{equation}
Similarly, we can write the SINR at the MN when the latter points its ULA to the $n$-th
IRS while eavesdropping the data stream intended for UE $k$, as
\begin{equation}
{\rm SINR}^{{\rm MN}}_{n,k,c} = \frac{\pi_k|\widetilde{\bv}_{n,c}\Herm \gammav_{k,c}|^2}
{\sum_{h\neq k}\pi_h|\widetilde{\bv}_{n,c}\Herm \gammav_{h,c}|^2+N_0B}
\end{equation}
where $\gammav_{k,c}$ is the $k$-th column of $\Gammam_c$ whose
expression is given in~\Eq{Gamma}, and $\widetilde{\bv}_{n,c}\Herm$ is defined in~\Eq{o}.

The data rate for UE $k$ under the $c$-th configuration can be computed as 
\begin{equation}
R(k,c)=B\log_2\left(1+{\rm SINR}^{{\rm UE}}_{k,c}\right)\,.
\end{equation}
Finally, the secrecy
rate (SR) obtained when the MN eavesdrops the data stream intended for UE $k$, by
pointing its antenna to the  IRS $n$, under configuration $c$, is given by:
\begin{equation}
  {\rm SR}(n,k,c) = \max      \big\{ 0, R(k,c) \mathord{-}  B\log_2(1\mathord{+}{\rm SINR}^{{\rm MN}}_{n,k,c})  \big\}\,.
\label{eq:SR}
\end{equation}
The $\max$ operator in~\Eq{SR} is required since, under certain
circumstances, ${\rm SINR}^{{\rm MN}}_{n,k,c}$ might be larger than
${\rm SINR}^{{\rm UE}}_{k,c}$.

\subsection{Communication scheme}
\label{sec:sub-time}
 
 Let us normalize time quantities  to the time it takes to receivers (legitimate or not) 
to switch from one configuration to another, and call such time interval {\em time unit}. 

Given~$\Cc$, the BS
chooses a set~$\bar{\Cc} \subseteq \Cc$ of configurations to activate, as well as a
criterion that legitimate users shall follow to determine the next
configuration to move to. In other words, legitimate nodes will always
know the next configuration to use while the eavesdropper cannot. 
A simple way to achieve this is to use {\em hash chains}~\cite{hussain2009key}: the first element of the chain 
(i.e., the first configuration to activate) is a secret shared among all legitimate nodes. Then, subsequent elements of the 
chain -- hence, subsequent configurations -- are achieved by hashing the current  element, 
in a way that is easy for honest nodes to compute, but impossible for an outsider to guess.
 We further assume that all chosen configurations are used with equal
probability, and that they are notified to legitimate users in a
secure manner, while the eavesdropper has no way of knowing the next
configuration in advance. As noted earlier, hash chains allow us to attain
both goals.

The decision about whether or not to use configuration~$c$ is expressed through
binary variables~$y(c)$, which take 1 if $c$ is adopted and 0 otherwise. 
Given the value of the decision variables~$y(c)$, 
we can write the probability~$\omega(k,n)$ that user~$k$ is served through IRS~$n$ under {\em any} 
of the chosen configurations $c\in \bar{\Cc}$, as
\begin{equation}
\omega(k,n)=\frac{\sum_{c\in\bar{\Cc}}\ind{\nu_c(k)=n}}{\left|\bar{\Cc}\right|}\,.
\end{equation}

As for the eavesdropper, we consider the most unfavorable scenario for the legitimate users 
and assume that 
the MN has already estimated the probability with which its victim is served by each IRS, and that it can 
leverage such information
by pointing its own beam towards each IRS according to those probabilities.

Given $\bar{\Cc}$, the BS sets 
 the number of time units $\tau \geq 1$ for which the legitimate users should stay with any  
configuration $c\in \bar{\Cc}$. 
Then, considering the fact  that
one time unit is the time needed to switching configuration and the communication
is paused during such switching time (i.e., every~$\tau+1$), it follows that the {\em
  average} rate for each legitimate user~$k$ can be written as:
\begin{equation}
\label{eq:rate-avg}
R_\text{avg}(k)=\frac{\tau}{\tau+1}\frac{\sum_{c\in\bar{\Cc}}R(k,c)}{\left|\bar{\Cc}\right|}\,.
\end{equation}

Moving to the eavesdropper, its objective is to have the smallest possible secrecy rate (SR) 
for its victim~$k^\star$. There are two strategies it can follow towards this end:
\begin{itemize}
    \item {\em static}: to always point towards the IRS that is most frequently used 
    to serve the victim~$k^\star$, i.e., \\
    $n^\star=\arg\max_n\omega(k^\star,n)$, or
    \item {\em dynamic}: to spend $\delta$ time units to try all IRSs, identify the one serving 
    the victim~$k^\star$, and then point towards it.
\end{itemize}
In the first case, the resulting SR is given by:
\begin{equation}
\label{eq:sr-static}
\text{SR}_\text{avg}^\text{static}(k^\star)=\frac{1}{\left|\bar{\Cc}\right|}
\sum_{c\in\bar{\Cc}}\text{SR}(n^\star,k^\star,c)\,,
\end{equation}
while in the latter case, the SR is as follows:
\begin{equation}
\label{eq:sr-dynamic}
\text{SR}_\text{avg}^\text{dynamic}(k^\star)=
\begin{cases}
\sum_{c\in\bar{\Cc}}\left[\frac{\tau\mathord{-}\delta}{\tau}\min_n\text{SR}(n, k^\star,c)
  \mathord{+}\frac{\delta R(k^\star,c)}{\tau} \right] \\ \quad\qquad\qquad\qquad\,\text{if~$\delta\le\tau$} \\
  \frac{1}{\left|\bar{\Cc}\right|}R(k^\star,c)\qquad \,\,\,\text{else}.
  \end{cases}
\end{equation}

The quantity within square brackets in \Eq{sr-dynamic} comes from the
fact that, for each configuration (i.e., each $\tau$~time units), the
eavesdropper spends~$\delta$ units trying all IRSs (during which the
SR will be~$R(k^\star,c)$, i.e., complete secrecy), and $\tau-\delta$ 
units experiencing the minimum secrecy rate across
all IRSs. 
In both cases, SR values are {\em subordinate to the fact that the
  BS is transmitting} -- clearly, if there is no
transmission, there can be no secrecy rate. Also, notice how we must
write SR values as dependent upon the eavesdropping victim~$k^\star$; 
indeed, the eavesdropper knows who its victim is, while legitimate
users do not.

The eavesdropper will choose the strategy that best suits it, i.e., results in the lowest secrecy rate. 
It follows that the resulting secrecy rate is:
\begin{equation}
\nonumber
\text{SR}_\text{avg}(k^\star)=
\min\left\{\text{SR}_\text{avg}^\text{static}(k^\star),\text{SR}_\text{avg}^\text{dyn}(k^\star)\right\}\,.
\end{equation} 

\section{Problem Formulation and Solution Strategy}
\label{sec:algo}

In this section, we first formulate the choice of set~$\bar{\Cc}\subseteq\Cc$ of configurations 
to enable as an optimization problem. 
Then, in light of the problem complexity, 
we propose an efficient heuristic called {\sf ParallelSlide}, and we show that 
the proposed algorithm obtains solutions provably 
close to the optimum in a remarkably short time.

\subsection{Problem formulation}
The goal of the network system  is to  maximize the average secrecy rate over time, 
so long as all legitimate users get at least an average rate~$R_{\min}$, i.e.,
\begin{equation}
\label{eq:obj}
\max_{\bar{\Cc},\tau}\min_{k}\text{SR}_\text{avg}(k)
\end{equation}
\begin{equation}
\label{eq:constr}
\text{s.t.}\quad R_{\text{avg}}(k)\geq R_{\min},\quad\forall k\,.
\end{equation} 
Notice  how objective \Eq{obj} must be stated in a max-min form: 
since the BS does not know who the eavesdropping victim is, 
it aims at maximizing the secrecy rate in the worst-case scenario, in which the node with the lowest SR 
is indeed the victim.

Furthermore, we remark that, in some cases, it may be necessary to use the 
same configuration~$c\in \Cc$ multiple 
times before repeating the cycle, i.e., to {\em replicate} a selected configuration.
Our system model and  notation  
 do not {\em directly} support this, as the decisions about 
configuration activation are binary (or, equivalently, a configuration cannot appear 
in set~$\bar{\Cc}$ more than once). However, it is possible to obtain the same 
effect as  repeating a configuration, by including several {\em replicas} thereof in~$\bar{\Cc}$: 
in this case, the data and secrecy rates of each replica of the configuration are 
evaluated separately, hence, the same configuration can be used multiple times if appropriate.

Next, to streamline the notation, let us indicate
with $\hat{R}(c)=\min_k R(k,c)$ the worst-case rate experienced by a
legitimate user under configuration~$c\in\Cc$, and with
$\hat{S}(c)=\min_n SR(n,k^\star,c)$ the minimum secrecy rate
experienced by the victim~$k^\star$ under such a configuration.
Importantly, secrecy rate values are averaged over (in principle) all possible positions of the 
eavesdropper, hence, computing such information requires {\em no knowledge} of the eavesdropper's 
position or channel quality. 
Then, let us assume  that the attacker follows the dynamic strategy, 
which has been proven~\cite{noi-wiopt21} to be the most effective except for very swift configuration changes.
By recalling that each configuration $c \in \bar{\Cc}$ is held for the same time duration, 
hence the temporal and the  numerical average coincide, 
we can rewrite \Eq{obj} as: 
\begin{eqnarray}
\nonumber \max_{\{y(c)\}_c} & \frac{\delta}{\tau+1}\frac{1}{\sum_{c \in \Cc} y(c)}\sum_{c\in\Cc}y(c)\hat{R}(c)+\\
& +\frac{\tau-\delta}{\tau+1}\frac{1}{\sum_{c \in \Cc} y(c)}\sum_{c\in\Cc}y(c)\hat{S}(c) \label{eq:obj2}\\
\mathrm{s.t.} & \frac{\tau}{\tau+1}\frac{1}{\sum_{c \in \Cc} y(c)}\sum_{c\in\Cc}y(c)\hat{R}(c){\geq} R_{\min}  \,.
\label{eq:constr}
\end{eqnarray}
The above expression accounts
for the fact that, within each time interval, legitimate users enjoy a 
secrecy rate equal to the average rate of the selected configurations for a
 fraction~$\frac{\delta}{\tau+1}$ of the time (during which the eavesdropper can hear nothing, hence, 
 the secrecy rate is the same as 
the UEs' data rate).
For the rest of the time, the secrecy rate is the average of the secrecy rates of the selected 
configurations. Constraint \Eq{constr} describes the fact that the system transmits nothing for 
a fraction~$\frac{1}{\tau+1}$ of the time, and legitimate users enjoy the average of the rates of 
the selected configurations for the rest of the time.

At last, notice that the problem above is combinatorial and nonlinear; hence, 
it is critical to envision a low-complexity heuristics that can cope with non-trivial instances of 
the problem while  yielding effective solutions.

\subsection{Solution strategy: The {\sf ParallelSlide} algorithm}

The NP-hardness of optimizing objective \Eq{obj2} subject to constraint \Eq{constr} means that making optimal decisions takes a 
prohibitively long time -- possibly, hours or days -- 
even for modestly-sized problem instances. 
We therefore opt for a heuristic approach, seeking to make high-quality -- 
namely, near-optimal decisions -- with small computational complexity, hence, in a short time.

Our high-level goal is to leverage the results of~\cite{wolsey1982maximising}, 
providing very good competitive ratio properties for a simple greedy algorithm, as long as (i) 
the objective is submodular nondecreasing, and (ii) the constraint is knapsack-like, 
i.e., additive. We will proceed as follows:
\begin{enumerate}
    \item Discussing the submodularity of the objective in \Eq{obj2} and of the constraint in \Eq{constr}, 
    showing that they are not submodular in general;
    \item Observing that, if the number~$|\bar{\Cc}|$ of configurations to eventually select were known, 
    then \Eq{obj2} and \Eq{constr} would be submodular;
    \item Exploiting the latter to propose an efficient algorithm solving the original problem.
\end{enumerate}

{\bf Submodularity.}
Recall that a generic set function $f(X)$ is submodular if, for any set~$\Ac$ and~$\Bc$ and element~$x$, the following holds:
\begin{equation}
\label{eq:submodular}
f(\Ac\cup\Bc\cup \{x\})-f(\Ac\cup\Bc)\leq f(\Ac\cup \{x\})-f(\Ac).
\end{equation}
Intuitively, adding~$x$ to a larger set~$\Ac\cup \Bc$ brings a lower benefit than adding it to a smaller set~$\Ac$; 
such an effect is often referred to as ``diminishing returns''.

Owing to its simplicity, let us focus on constraint \Eq{constr} and derive a restrictive, 
necessary (and sufficient) condition for its submodularity.
\begin{property}
\label{prope:restrictive}
Constraint \Eq{constr} is submodular only if configurations are selected from the worst-performing 
to the best-performing one.
\end{property}
\begin{proof}
We start from the submodular definition in \Eq{submodular}, where in our case~$\Ac\subseteq\Cc$ 
and~$\Bc\subseteq\Cc$ are sets of configurations. Let~$A$ and~$B$ denote their cardinality, 
and~$a$ and~$b$ define their corresponding average data rate. Furthermore, 
let $\rho=\hat{R}(c')$~be the rate of the new configuration~$c'\in\Cc$.
Keeping in mind that the $\frac{\tau}{\tau+1}$~terms simplify away, \Eq{submodular} becomes:
\begin{equation}
\frac{Aa{+}Bb{+}\rho}{A{+}B{+}1}{-}\frac{Aa{+}Bb}{A{+}B}
\leq \frac{Aa{+}\rho}{A{+}1} {-}\frac{Aa}{A},
\end{equation}
which simplifies to
\begin{equation}\label{eq:AB-ineq}
a (2 A + B + 1)\leq(A + 1) b + \rho (A + B),
\end{equation}
Notice that (\ref{eq:AB-ineq}) holds if~$\rho\geq b\geq a$, as per the hypothesis.
\end{proof}
The condition derived in \Prope{restrictive} is very restrictive; indeed, 
there is no good reason why the worst-performing configurations should be chosen {\em first}. 
Also notice that the non-submodularity of objective \Eq{obj2} can be proven through the very same argument.
\begin{figure}
\centering
\includegraphics[width=1.0\columnwidth]{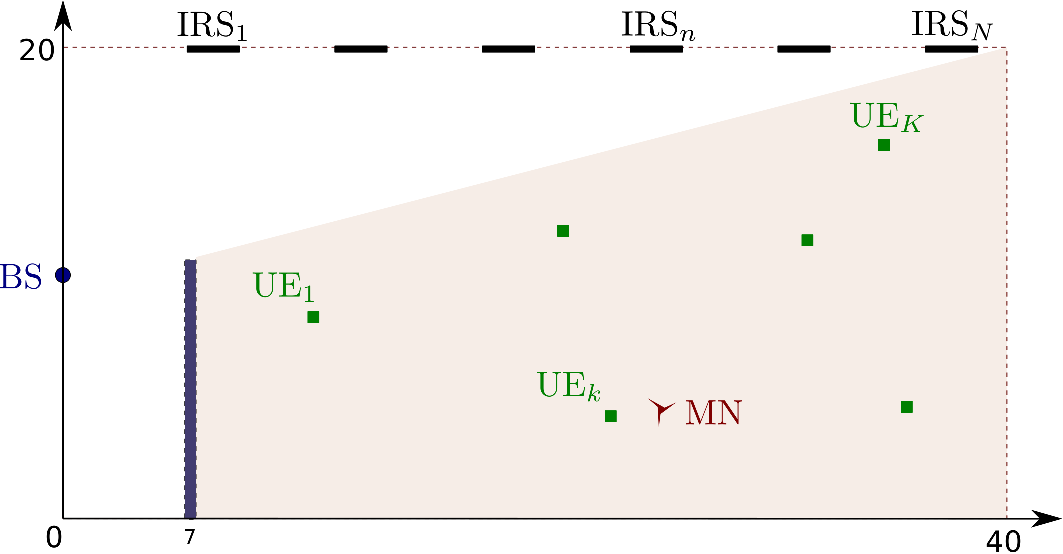}
\caption{Base scenario.\label{fig:scenario}}
\end{figure}

\begin{figure*}
\centering
\includegraphics[width=.8\columnwidth]{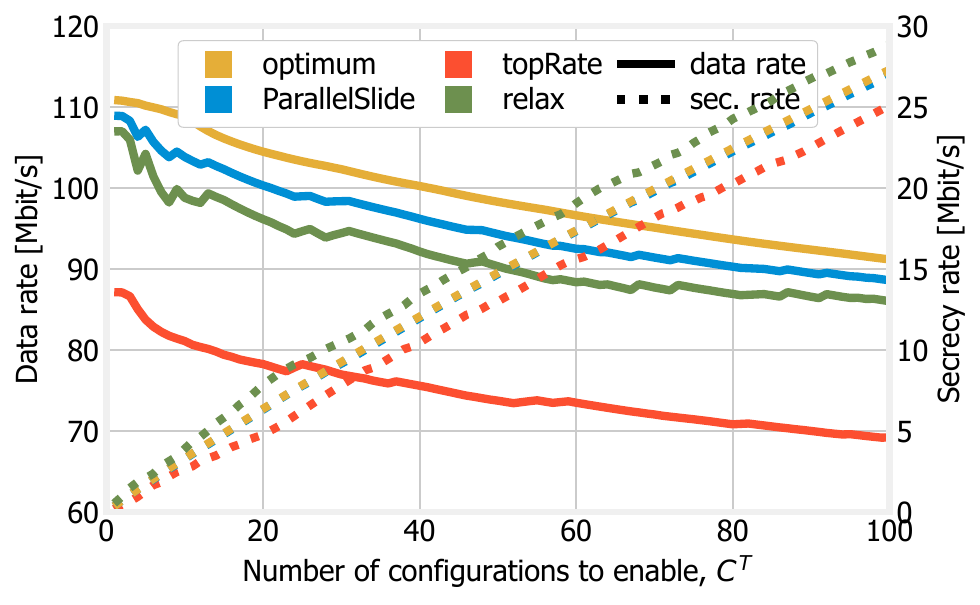}
\includegraphics[width=.8\columnwidth]{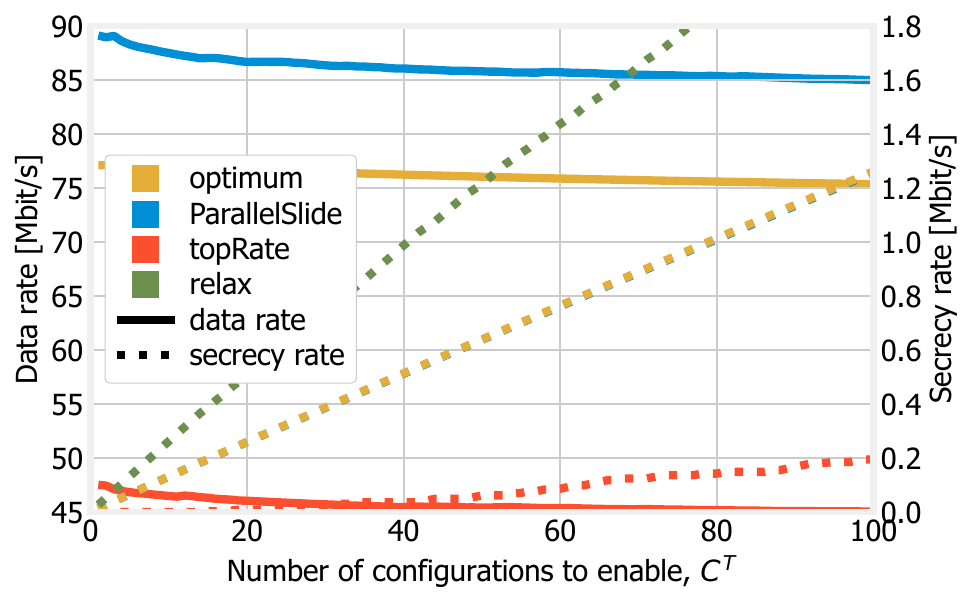}
\caption{
Data rate and secrecy rate achieved as the number~$C^\text{T}$ of configurations to choose changes, 
under the base (left) and extended (right) scenarios.
    \label{fig:general}
} 
\end{figure*}

{\bf Adding an oracle: the {\sf ParallelSlide} algorithm.}
Intuitively, what destroys the submodularity of \Eq{obj2}--\Eq{constr} is the presence of the average, 
which in turn comes from the fact that we must choose both {\em how many} configurations to select, 
and {\em which} ones. Splitting the two parts of the problem would indeed result in a significantly 
better-behaved problem.
More specifically, we can prove that the following property holds.
\begin{property}
\label{prope:easy}
If the number~$|\bar{\Cc}|$ of configurations to choose is known, then objective \Eq{obj2} is submodular, 
and constraint \Eq{constr} is a knapsack constraint.
\end{property}
\begin{proof}
Concerning the objective, the proof trivially comes from the observation that, 
once $|\bar{\Cc}|$~is known, \Eq{obj2} reduces to a sum of (i) constant quantities, and (ii) 
decision variables multiplied by positive coefficients.

As for constraint \Eq{constr}, recall that a knapsack constraint 
over set~$\Vc$ is  an alternate to a cardinality constraint where each element of the set has a cost 
and the selected items cannot exceed a total budget~\cite{tang2021revisiting}.  
We can re-write \Eq{constr} as:
\begin{equation}
\nonumber
\sum_{c\in\Cc}y(c)\hat{R}(c)\geq \frac{\tau}{\tau+1} R_{\min},
\end{equation}
hence, it is a knapsack constraint.
\end{proof}

\Prope{easy} implies that, once~$|\bar{\Cc}|$ is known, the problem reduces to optimizing a 
submodular nondecreasing function subject to a knapsack constraint. 
Such problems can be solved very efficiently by greedy 
algorithms~\cite{wolsey1982maximising,tang2021revisiting}, picking at each step  
the configuration with the highest benefit-to-cost ratio. 
We leverage this principle while designing our algorithm, called {\sf ParallelSlide} 
and presented in \Alg{parslide}. The basic idea of {\sf ParallelSlide} is indeed to (i) 
try all possible sizes of~$\bar{\Cc}$, 
and (ii) for each target size, obtain a solution 
with that size by applying the benefit-to-cost ratio principle 
of~\cite{wolsey1982maximising,tang2021revisiting}. 
We remark that the number of possible sizes of $\bar{\Cc}$ cannot exceed 
the minimum between the number of all possible configurations 
and the maximum number of configurations that it is possible to select, i.e.,~$|\Cc|$.

Specifically, 
for each value of the target size~$C^\text{T}$ (\Line{forall-c} in \Alg{parslide}), 
the algorithm builds a solution by first using all configurations (\Line{use-all}), 
and then removing, at each iteration, the configuration which minimizes 
the ratio between the data rate it yields  and 
the corresponding  secrecy rate, as per \Line{ratio}. In \Line{ratio},~$\hat{R}$ and~$\hat{S}$ indicate, 
respectively, the rate and secrecy rate obtained after activating configuration~$c$, 
accounting for the fact that a total of $C^\text{T}$~configurations will eventually be activated.

Upon reaching size~$C^\text{T}$, the algorithm checks if set $\text{\path{used\_configs}}$ 
results in a feasible solution (\Line{check-feasible}) and, if so, adds it to the 
set~$\text{\path{feasible\_solutions}}$ (\Line{add-feasible}). After trying out all possible values 
of~$C^\text{T}$, the solution resulting in the best performance, i.e., the largest value 
of objective \Eq{obj2}, is selected.

\subsection{Algorithm analysis}
The {\sf ParallelSlide} algorithm has two very good properties, 
namely (i) it has a very low computational complexity, and (ii) it provides results that are 
provably close to the optimum. 
Let us begin from the former result. 
\begin{property}
The {\sf ParallelSlide} algorithm has quadratic worst-case computational complexity, namely,~$O(|\Cc|^2)$.
\label{prope:quadratic}
\end{property}
\begin{proof}
The proof comes by inspection of \Alg{parslide}. The algorithm contains two loops: 
an outer one (beginning in \Line{forall-c} that runs exactly~$|\Cc|$ times), and 
an inner one (beginning in \Line{c-target} that runs at most~$|\Cc|$ times). 
All other operations, e.g., checking feasibility in \Line{check-feasible}, are 
elementary and are run fewer than~$|\Cc|^2$ times. Hence, the final worst-case 
computational complexity is~$O(|\Cc|^2)$.
\end{proof}
Its quadratic complexity allows {\sf ParallelSlide} to make swift decisions, and makes 
it suitable for real-time usage.

\begin{algorithm}
\caption{The {\sf ParallelSlide} algorithm
\label{alg:parslide}
} 
\begin{algorithmic}[1]
\Require{$\Cc$}
\State{$\text{\path{feasible\_solutions}}\gets\emptyset$}
\ForAll{$C^\text{T}\in[|\Cc|,\dots,1]$} \label{line:forall-c}
 \State{$\text{\path{used\_configs}}\gets\Cc$} \label{line:use-all}
 \While{$|\text{\path{used\_configs}}|>C^\text{T}$} \label{line:c-target}
  \State{$\text{\path{to\_del}}\gets\arg\min_{c\in\text{\path{used\_configs}}}\frac{\mathbf{\hat{R}}(c,C^\text{T})}{\mathbf{\hat{S}}(c,C^\text{T})}$} \label{line:ratio}
  \State{$\text{\path{used\_configs}}\gets\text{\path{used\_configs}}\setminus\{\text{\path{to\_del}}\}$}
 \EndWhile
 \If{$\mathbf{is\_feasible}(\text{\path{used\_configs}})$} \label{line:check-feasible}
  \State{$\text{\path{feasible\_solutions}}\gets$}\label{line:add-feasible}
  \State*{$\text{\path{feasible\_solutions}}\cup\{\text{\path{used\_configs}}\}$} 
 \EndIf
\EndFor
\Return{$\arg\max_{\text{\path{s}}\in\text{\path{feasible\_solutions}}}\mathbf{secrecy\_rate}(s)$} \label{line:return}
\end{algorithmic}
\end{algorithm}

Concerning the quality of decisions, we are able to prove that:
\begin{itemize}
\item {\sf ParallelSlide} is remarkably close to the optimum, and
\item the distance between {\sf ParallelSlide} and the optimum does not depend upon the problem size.
\end{itemize}
More formally, the ratio of the objective value \Eq{obj2} obtained by {\sf ParallelSlide} 
to the optimal one is called {\em competitive ratio}. In most cases, competitive ratios decrease 
(i.e., the solutions get worse) as the problem size increases; intuitively, 
larger problems are harder to solve. This is not the case of {\sf ParallelSlide}, 
whose competitive ratio is constant, as per the following property:
\begin{property}
The {\sf ParallelSlide} algorithm has a constant competitive ratio of~$0.405$.
\label{prope:competitive}
\end{property}
\begin{proof}
The proof comes from observing that the inner loop of \Alg{parslide}, i.e., 
the one starting at \Line{forall-c}, mimics the \path{MGreedy} algorithm presented 
in~\cite[Alg.~1]{tang2021revisiting}. Our problem has one additional potential 
source of suboptimality, namely, the choice of the number~$C^\text{T}$ of configurations 
to choose; however, the outer loop of \Alg{parslide} tries out all possible values 
of~$C^\text{T}$ (\Line{forall-c}) and chooses the one resulting in the best performance 
(\Line{return}). It follows that no further suboptimality is introduced, and {\sf ParallelSlide} 
has the same competitive ratio as~\cite[Alg.~1]{tang2021revisiting}, namely,~$0.405$.
\end{proof}

\begin{figure*}
\centering
\includegraphics[width=.8\columnwidth]{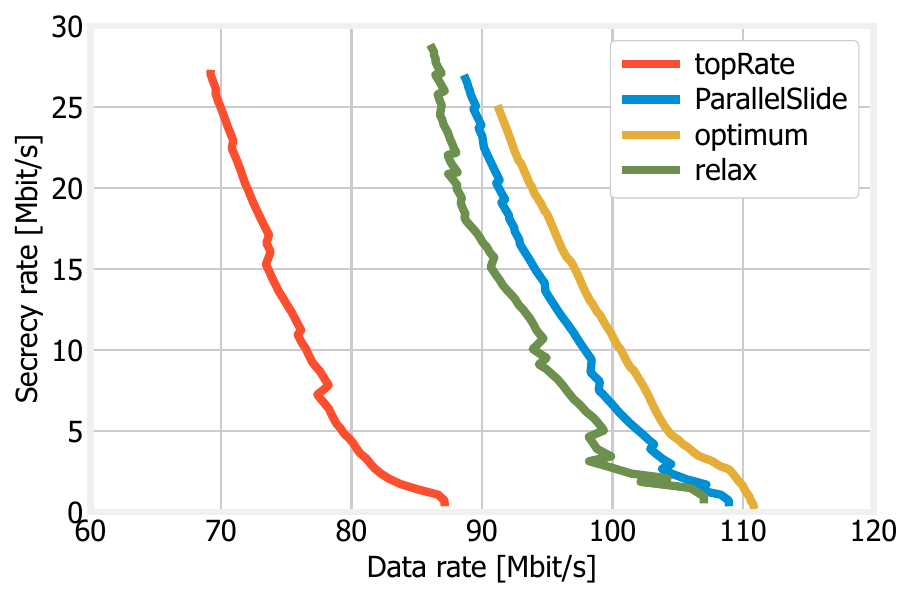}
\includegraphics[width=.8\columnwidth]{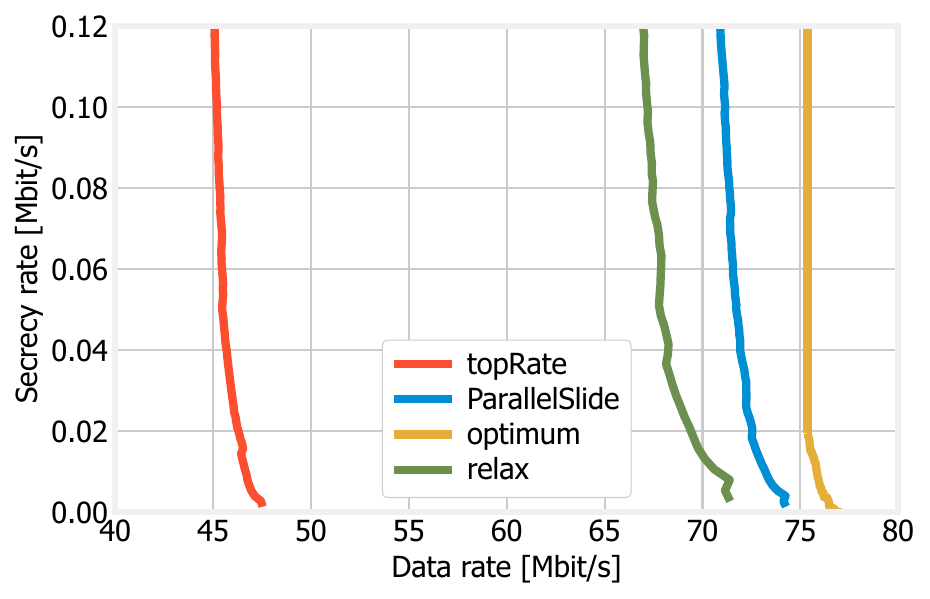}
\caption{
Trade-offs between data rate and secrecy rate attained by different strategies, under the base (left) and extended (right) scenarios.
    \label{fig:trajectories}
} 
\centering
\subfigure[]{\includegraphics[width=.31\textwidth]{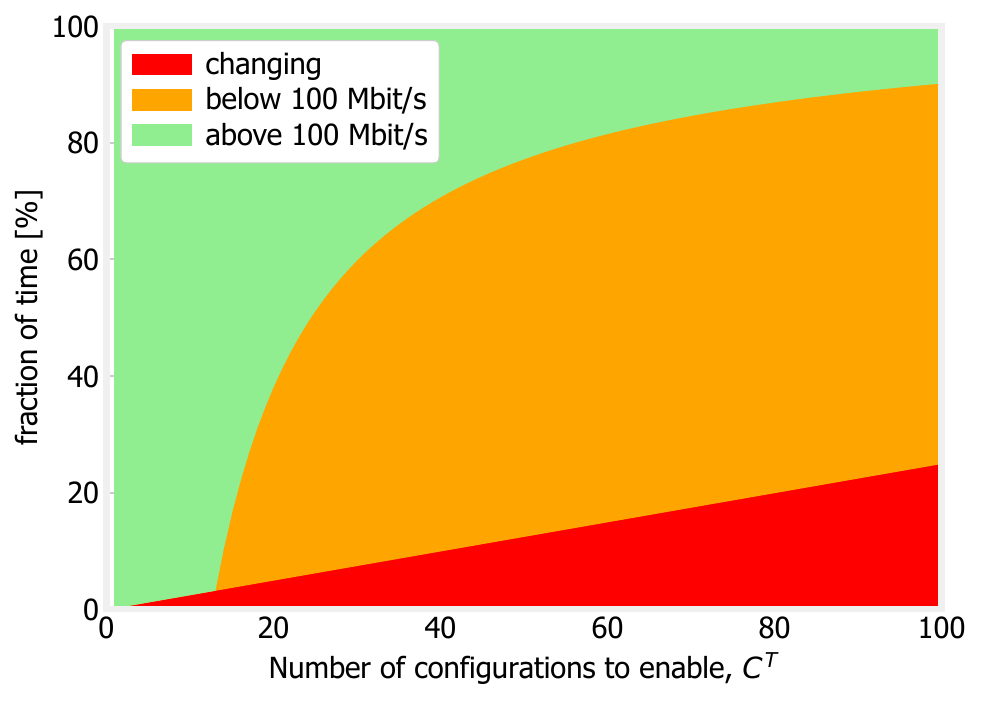}}
\subfigure[]{\includegraphics[width=.31\textwidth]{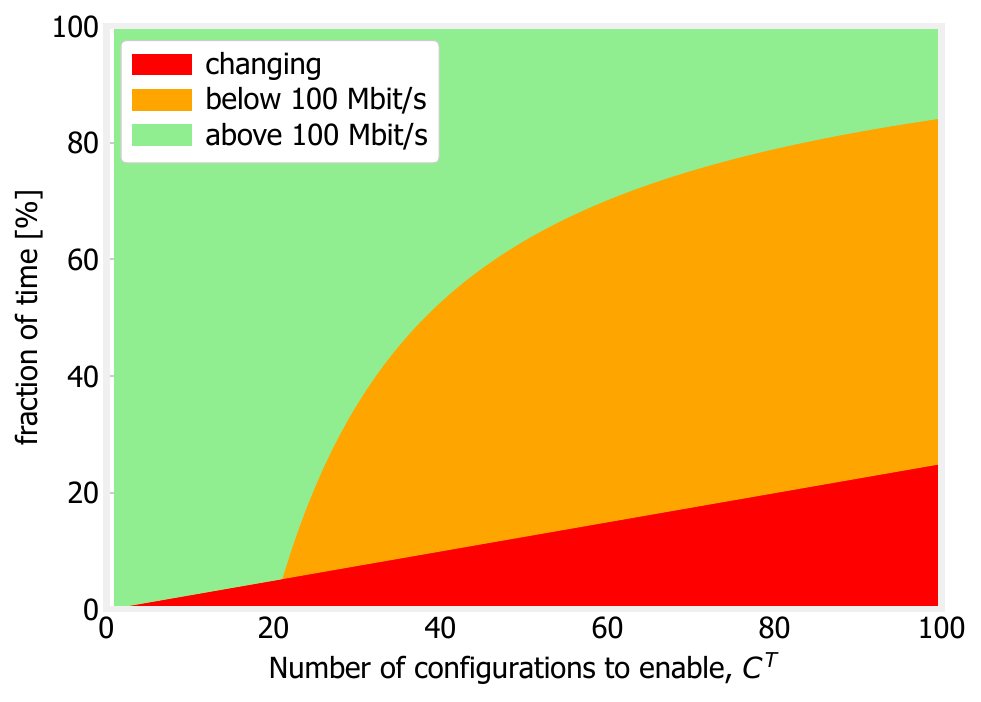}}
\subfigure[]{\includegraphics[width=.31\textwidth]{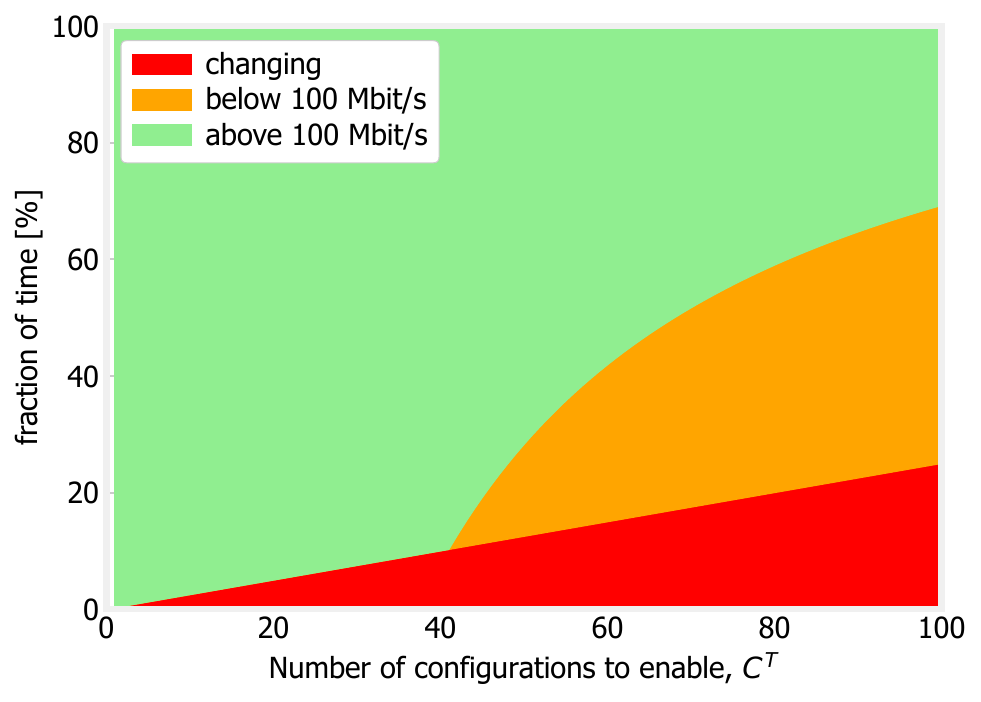}}
\caption{
Base scenario: fraction of time spent by nodes enacting higher-data rate configurations (green), 
enacting lower-data rate configurations (orange), or switching between configurations (red), 
under the topRate (a), {\sf ParallelSlide} (b), and optimum (c) strategies.
    \label{fig:filled}
} 
\end{figure*}

Finally, we can prove that {\sf ParallelSlide} does in fact convergence after a finite number of iterations:
\begin{property}
The {\sf ParallelSlide} algorithm converges after at most~$|\Cc|^2$ iterations. 
\end{property}
\begin{proof}
The proof comes by the inspection of \Alg{parslide}, which has two nested loops, each of which runs at most~$|\Cc|$ times.
\end{proof}

So far, we have presented and discussed {\sf ParallelSlide} with reference to a scenario where no LoS path from the BS to any user exists. These are indeed the most challenging scenarios, 
and those where IRSs are most useful; however, {\sf ParallelSlide} works unmodified when direct paths do exist. Specifically:
\begin{itemize}
    \item the set of IRSs is extended with an extra item~$\varnothing$, indicating that the direct path is used;
    \item additional configurations are generated accordingly;
    \item {\sf ParallelSlide} is applied to the new set of configurations, with no change.
\end{itemize}

\section{Performance Evaluation}
\label{sec:peva}
To study the performance of ParallelSlide, we consider a
  scenario where BS, UEs, IRSs and the MN are located in a room of
  size 40\,m$\times$20\,m (see \Fig{scenario} for details).
  As can be observed, the BS--UEs LoS path is unavailable since it is
  blocked by an obstacle, which is   
   the most challenging scenario for our 
  decision-making process.

  The network operates
in the sub-terahertz spectrum, namely,
  at central frequency $f_c\mathord{=}100$\,GHz,
  corresponding to the wavelength $\lambda{=}3$\,mm. The BS, whose ULA
  is composed of $M_{\rm BS}{=}32$ isotropic (0\,dBi) antenna elements,
  is located at coordinates $(0,10)$\,m; the BS antenna gain is,
  thus, $G{=}M_{\rm BS}$.  The signal bandwidth is $B{=}1$\,GHz, and the
  transmit power is set to $P_t{=}10$\,dBm. Such power is equally shared
  among UEs, i.e., the matrix $\Pim$ in~\Eq{Gamma} is
  proportional to the identity matrix.
    
    In our scenario all $N$ IRSs are identical, have square shape, and are made
    of 128$\times$128 meta-atoms, (i.e., $L_n{=}128$, $n{=}1,\ldots,N$)
    with no gaps between them. We also consider that meta-atoms have square
    shape with side length $\lambda/2$, so that each IRS has area $A=L^2
    \lambda^2/4{=}368.64$\,cm$^2$.  Also, IRSs are placed on the topmost
    wall and equally spaced.
    
  The UEs are uniformly distributed in the shaded area shown in
  \Fig{scenario}. All UEs are equipped with ULAs, each composed of $M_{\rm UE}{=}8$
  isotropic (0\,dBi) antenna elements, hence their antenna gain is
  $G{=}M_{\rm UE}$.    
    The malicious node too is equipped with $M_{\rm MN}=8$ isotropic
    antenna elements and is randomly located around the eavesdropped
    UE.

In order to study the performance of ParallelSlide in the most challenging conditions,
the position of the malicious node is uniformly distributed in a 
    square of side 1\,m around the eavesdropped UE.  Finally, at both
    UEs and MN receivers, the noise power spectral density is set to
    $N_0=-174$\,dBm/Hz.

    Specifically, we consider the following two simple, yet representative, cases:
\begin{itemize}
    \item a {\bf base} scenario, including a total of $K{=}6$~legitimate users and $N{=}6$~IRSs, 
    resulting in a total of $|\Cc|=6!{=}720$~possible configurations;
    \item an {\bf extended} scenario, where we increase the number of users and IRSs 
    to~$N{=}K{=}8$ (hence, the number of possible configurations grows to $8!\mathord{=}40\mathord{,}320$).
\end{itemize}

We compare {\sf ParallelSlide} against three alternative solutions, namely:
\begin{itemize}
    \item A simple {\em topRate} approach, selecting the $C^\text{T}$ configurations with the highest rate;
    \item A strategy, labelled {\em relax} in plots, and performing a relaxation of the problem to solve as per~\cite{8359418};
    \item The {\em optimum}, found through simulated annealing.
\end{itemize}
The ``relax'' strategy follows the strategy of~\cite{8359418}, and performs the following main operations:
\begin{enumerate}
    \item It solves an LP (linear problem) relaxation of the problem in \Eq{obj}, where binary variables~$y(c)$ are replaced by real ones $\bar{y}\in[0,1]$;
    \item It incrementally activates more configurations, choosing them with a probability proportional to~$\bar{y}(c)$;
    \item It stops upon reaching the target number of configurations.
\end{enumerate}
For simulated annealing, we use the following parameters:
\begin{itemize}
    \item number of generations: 50;
    \item solutions per population 100;
    \item parents mating: 4;
    \item mutation probability: 15\%;
    \item crossover type: single point;
    \item gene space: $\{0, 1\}$;
    \item number of genes: $|\Cc|$.
\end{itemize}

\Fig{general} depicts how the number~$C^\text{T}$ of configurations to choose influences the resulting 
rate and secrecy rate, under {\sf ParallelSlide} and its counterparts.
As it can be expected, the achievable rate (left-hand side scale) is always substantially 
higher than the secrecy rate (right-hand side scale). 
The goal of our performance evaluation is not to directly compare the two metrics; rather, 
we evaluate how different strategies (corresponding to different colors in the plots) 
impact both metrics (represented by different line styles in the plots).

A first important observation 
we can make is that solid and dotted curves in the plots, representing (respectively) data 
rate and secrecy rate, have different slopes. Specifically, choosing more configurations decreases 
the data rate, as we are forced to include lower-rate IRS-UE assignments. On the other hand, 
more configurations result in a better secrecy rate, as it takes longer for the eavesdropper 
to guess the configuration adopted by the legitimate nodes.

Concerning the relationship between the strategies, we can observe that {\sf ParallelSlide} 
consistently and significantly outperforms both the ``topRate'' and ``relax'' benchmarks, and almost matches the optimum for all 
values of~$C^\text{T}$. This validates the intuition from which {\sf ParallelSlide} stems, 
i.e., combining both rate and secrecy rate when making configuration-selection decisions, results in better performance. 
It is also interesting to remark how {\sf ParallelSlide}'s performance is very close to the optimum, 
even more than foreseen by the bound in \Prope{competitive}.

\Fig{trajectories} offers additional insights on the different performance of {\sf ParallelSlide} 
and its alternatives, summarizing the trade-offs between data rate and secrecy rate they are able to 
attain. We can observe that {\sf ParallelSlide} can achieve higher-quality trade-offs; 
in other words, for a given value of minimum data rate ($R_{\min}$ in \Eq{constr}), {\sf ParallelSlide} 
can obtain a better secrecy rate, i.e., a higher value of the objective in \Eq{obj}.

In summary, we can conclude that {\sf ParallelSlide}'s ability to account for both data rate and 
secrecy rate when making configuration-selection decisions allows it to attain high-quality 
trade-offs between such two quantities, thus outperforming alternative approaches and 
closely matching the optimum.

We now focus on the base scenario, and seek to better understand the effect of adding more 
configurations, i.e., increasing~$C^{\text{T}}$. To this end, we plot in \Fig{filled} the 
fraction of time spent by nodes:
\begin{itemize}
    \item enacting higher-data rate configurations, resulting in a rate above 100~Mbit/s (green);
    \item enacting lower-data rate configurations, with a rate below 100~Mbit/s (orange);
    \item idle, switching between configurations (red).
\end{itemize}

We can observe that increasing~$C^{\text{T}}$ adversely impacts the rate (as per \Fig{general}) 
in two main ways. On the one hand, more time is spent switching between configurations, as switches 
themselves become more frequent. At the same time, selecting more configurations means, 
necessarily, selecting {\em slower} IRS-UE assignments, further decreasing the resulting average rate. 
Comparing the plots, we can observe that the performance difference between different strategies 
only comes from the ability to select better (i.e., higher-data rate) configurations, as the 
time spent switching between configurations only depends upon~$C^{\text{T}}$ and is not impacted 
by the strategy being used.

Overall, \Fig{filled} confirms our intuition that $C^{\text{T}}$~should only be as large as 
needed to attain the required secrecy rate, and further increasing it would needlessly hurt 
the performance.

\section{Conclusions}
\label{sec:concl}

We have addressed the issue of defending from passive eavesdropping in wireless networks 
powered by intelligent reflective surfaces (IRSs). After modeling such a scenario, 
we have identified the latent tension between the objective of guaranteeing a high data rate 
{\em and} a high secrecy rate to the legitimate network users.

Accordingly, we have proposed an efficient and effective decision-making strategy, 
called {\sf ParallelSlide}, that achieves high-quality trade-offs between data rate and secrecy rate. 
After proving that {\sf ParallelSlide} has a polynomial computational complexity and a constant 
competitive ratio, we have showed through numerical evaluation that it significantly 
outperforms alternative approaches, and closely matches the optimum.

\section*{Acknowledgment}
This work was supported by the Horizon Europe project CENTRIC (Grant No. 101096379).

\bibliographystyle{elsarticle-num}
\bibliography{refs}

\end{document}